\documentclass[10pt,twocolumn,letterpaper]{article}

\usepackage{cvpr}
\usepackage{times}
\usepackage{epsfig}
\usepackage{graphicx}
\usepackage{amsmath}
\usepackage{amssymb}

\usepackage[pagebackref=true,breaklinks=true,letterpaper=true,colorlinks,bookmarks=false]{hyperref}

\cvprfinalcopy 

\ifcvprfinal\fi
\begin{document}

\title{DRCAS: Deep Restoration Network for Hardware Based \\Compressive Acquisition Scheme}

\author{ \textbf{Pravir Singh Gupta\textsuperscript{1}, Xin Yuan\textsuperscript{2}, Gwan Seong Choi\textsuperscript{1}}\\ \\ 
	\large{\textbf{Texas A\&M University, College Station, Texas, USA\textsuperscript{1}}}\\ \large{\textbf{Nokia-Bell Labs, Murray Hill, New Jersey, USA\textsuperscript{2}}}\\
	\tt \small{pravir@tamu.edu, xyuan@bell-labs.com, gwanchoi@tamu.edu}}

\maketitle

\begin{abstract}
We investigate the power and performance improvement in image acquisition devices by the use of CAS (Compressed Acquisition Scheme) and DNN (Deep Neural Networks). Towards this end, we propose a novel image acquisition scheme HCAS (Hardware based Compressed Acquisition Scheme) using hardware-based binning (downsampling), bit truncation and JPEG compression and develop a deep learning based reconstruction network for images acquired using the same. HCAS is motivated by the fact that in-situ compression of raw data using binning and bit truncation results in reduction in data traffic and power in the entire downstream image processing pipeline and additional compression of processed data using JPEG will help in storage/transmission of images. The combination of in-situ compression with JPEG leads to high compression ratios, significant power savings with further advantages of image acquisition simplification. Bearing these concerns in mind, we propose DRCAS (Deep Restoration network for hardware based Compressed Acquisition Scheme), which to our best knowledge, is the first work proposed in the literature for restoration of images acquired using acquisition scheme like HCAS.  When compared with the CAS methods (bicubic downsampling) used in super resolution tasks in literature, HCAS proposed in this paper performs superior in terms of both compression ratio and being hardware friendly. The restoration network DRCAS also perform superior than state-of-the-art super resolution networks while being much smaller. Thus HCAS and DRCAS technique will enable us to design much simpler and power efficient image acquisition pipelines. 
\end{abstract}


\section{Introduction}
We are living in a multimedia world. With the advent of Internet and mobile devices, the amount of multimedia content generated by users is increasing at a tremendous rate. In addition, with the improvement in VLSI (Very Large Scale Integration) technology, resolution of image sensors are also increasing. Smartphones with image sensor resolution greater than 20 Megapixel are commonly used and image sensor vendors are also offering sensors more than 40 Megapixel range. However, lately Moore's law has started to saturate and hence there is an increasing pressure to extract performance improvements from architectural and algorithmic innovations than device scaling. Videos and images present a huge burden in processing, storage and transmission networks. It also presents a challenge in terms of power consumption in image acquisition devices such as mobile devices. 
\begin{figure}[tb]
	\centering
	\includegraphics[width=.5\textwidth]{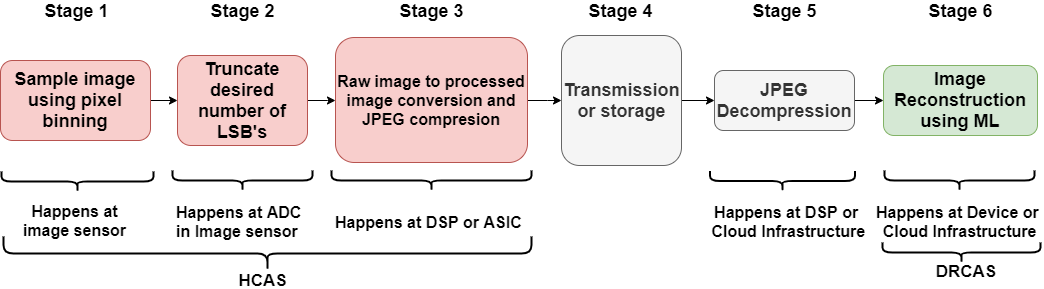}
	\caption{Proposed Image Acquisition Methodology:  It consists of 6 stages. In the stages 1-3 (HCAS), an image gets compressed using downsampling, bit truncation and JPEG. Stage 4 represents transmission of image which can be a wireless medium or an on-chip bus or even storage. Stage 5 performs JPEG decompression and stage 6 consists of the proposed DRCAS to restore the desired image.}
	\label{Fig:methodology}
\end{figure}
\begin{figure}[htbp!]
	\begin{center}
		\includegraphics[height=5cm]{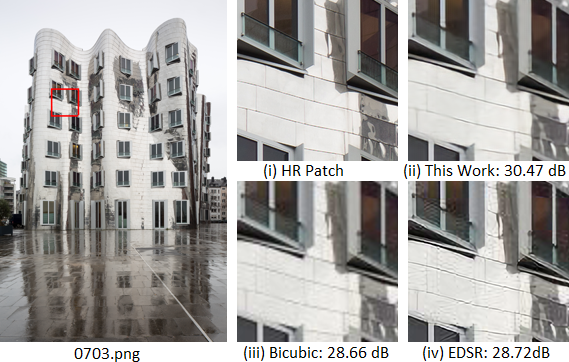}
	\end{center}
	\caption{Left: 0703 image from  the DIV2K  dataset \cite{div2k}.
		Downsampling is performed via 2$\times$2 binning, with bit truncation such that B.W. (bit width) = 7 and JPEG compression at quality $Q=~90$.
		Right: (i) HR patch from the left image, (ii) reconstruction using our proposed DRCAS, (iii) reconstruction using Bicubic interpolation, and (iv) reconstruction with EDSR~\cite{edsr}. }	
	{\label{Fig:comparison}}
\end{figure}

On the other hand, the advanced machine learning (ML) techniques especially deep learning perform much better than traditional computer vision techniques for tasks like super-resolution, object detection \etc 
Inspired by this, this paper exploits the power of deep learning techniques to reduce the data traffic in the {\em image acquisition pipeline} which will make it easier to meet power and performance requirements.
\cite{resnet,ntire}.  
{Motivated by the fact that if hardware constraints of imaging system are taken into consideration then deep neural networks (DNN) will be able to address these problems of imaging systems, we propose HCAS (Hardware based Compression Acquisition Scheme)  and a reconstruction network for the same. The proposed HRCAS consists of combination of binning (downsampling), bit truncation and JPEG compression. To reconstruct the original image, we propose DRCAS, which is a DNN based reconstruction network to perform image restoration by super-resolution (to restore loss of resolution caused by binning) and artifact removal (caused by bit truncation and JPEG compression).}


%
The proposed acquisition scheme HCAS is shown in Fig. \ref{Fig:methodology},
which is different from existing super-resolution networks,  
and one exemplar reconstruction is shown in Fig. \ref{Fig:comparison}. 

{
While DNN has been used for tasks like super-resolution (Ref. \cite{srcnn,accSR,srgan,wangSurvey,yangSurvey}) and image denoising \cite{denoising}, this work is novel 
in the following perspectives:                 
\begin{itemize}
	\item [i)] We propose a new image acquisition framework, HCAS, (Fig.~\ref{Fig:methodology}), which is compression based. It uses realistic and hardware based compression schemes from imaging system perspective like binning, bit truncation and JPEG compression. It performs compression on entire image acquisition pipeline i.e. from raw data at source (image sensor) to processed data (using JPEG). 
	\item [ii)]  Downsampling operation in our framework is {\em averaging and rounding} instead of bicubic which is more popular for superresolution tasks \cite{ntire}. Again our method is more realistic as averaging operation is easy to implement in hardware especially at image sensor level using pixel binning technique available in commercial image sensors.
	\item[iii)] By performing in-situ compression on raw data using binning and bit truncation, HCAS performs power savings in downstream power hungry components like ADC (Analog to Digital Converters) and DSP units.
\end{itemize}}


Following this, our proposed sensing framework HCAS in Fig.~\ref{Fig:methodology} results in significant reduction in raw data rate \ie data generated from image sensor as well as final image size. This has direct potential for power savings, transmission bandwidth savings and simplified hardware implementation. One can argue that instead of downsampling one can use a low resolution image sensor itself. This argument will hold valid only if reconstruction or super resolution process is exact, which is not possible. Thus downsampling using binning gives user an option to choose between the two depending upon requirements. There are some other works in the literature which have proposed pixel bit depth enhancements~ \cite{bitEnConv,becalf,bitEn}. However these works were targeted at converting low bit depth  images to high bit depth display. Another work proposed super-resolution and bitdepth enhancement~\cite{supBitEn}, however, the authors used A+ (Adjusted anchored neighborhood regression algorithm) \cite{a+}  method for super-resolution instead of Neural Networks. Some works have also focused on denoising and compression artifact removal tasks \cite{symSkip,artifactRed,denoising}.  To the best of our knowledge, this is the first work to investigate the DNN based image restoration considering the  combination of downsampling, bit truncation and JPEG for compression and power savings.

\section{Background}
In this section, we review the background of building blocks of the proposed image acquisition methodology in Fig.~\ref{Fig:methodology}.
Via introducing the image sensors in Sec.~\ref{Sec:imsensor}, we understand why {\em binning} is preferred in hardware. 
By analyzing the power consumption in Sec.~\ref{Sec:pwrPerf}, we understand how to save power of the device using super resolution. 
The JPEG compression is reviewed in Sec.~\ref{Sec:JPEG}, and following this, the transmission and storage techniques are described in Sec.~\ref{Sec:tranNstorage}. 
As a widely used interpolation approach, Bicubic interpolation is reviewed in Sec.~\ref{Sec:Bicubic} and the residual network, which will be used in our design is introduced in Sec.~\ref{Sec:ResNet}.

\subsection{Image sensors} \label{Sec:imsensor}
\begin{figure}[htb]
	\centering
	\vspace{-3mm}
	\centerline{\includegraphics[width=0.5\textwidth]{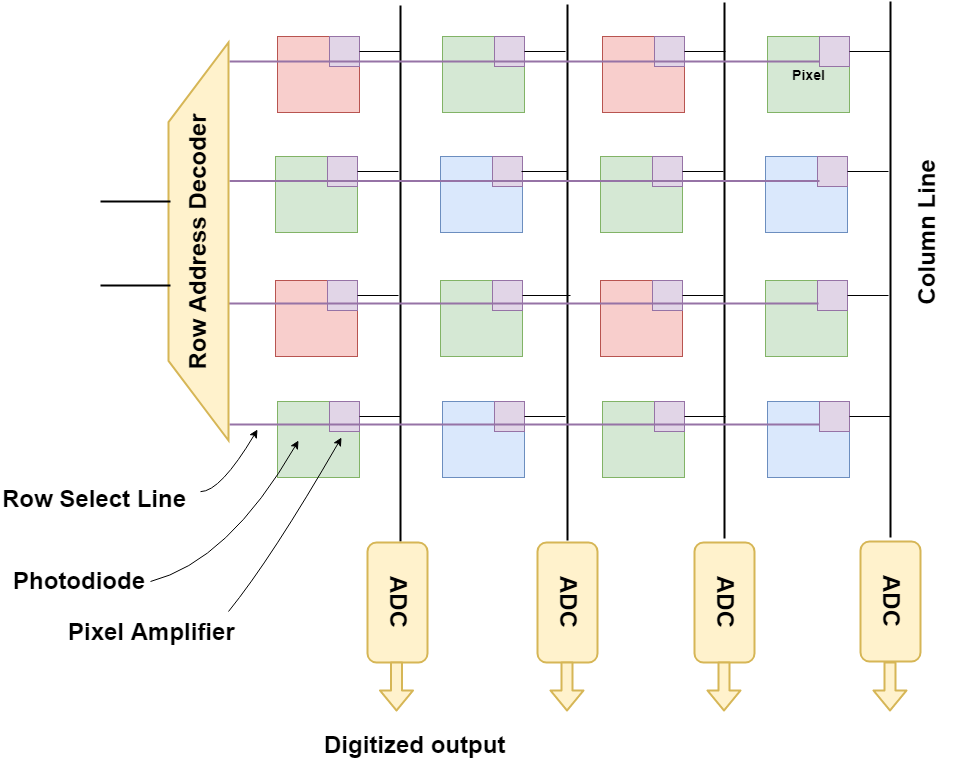}}
	\caption{A simple schematic of image sensor.}
	\label{Fig:IS}
\end{figure}
An image sensor converts light intensity into electrical signals. It consists of a rectangular grid of pixels. Typically pixels are addressed in a row wise fashion \ie all pixels in the same row are addressed simultaneously. The addressed row outputs the signal in column lines which is connected to Analog-to-Digital Converter (ADC). The ADC produces digitized version of the image. The number of digitized bits depend upon the resolution of image sensor. A simple schematic of image sensor is shown in Fig. \ref{Fig:IS}.

Most commercially available image sensors have a feature called "binning" which simply means to {\em combine information of a group of neighboring pixels together to form a single pixel in the output image}. Binning results in loss of spatial resolution by a factor equal to the number of pixels binned together, which can be either an addition operation or an averaging operation on image pixels. Addition operation has the advantage in low light situations as it increases the low light sensitivity of image sensor~\cite{lLBinning}; on the other hand, averaging operation has advantage in common lighting conditions as it prevents saturation of pixels in that scenario~\cite{frameMultires}. Binning has been used in various applications successfully like low light imaging \cite{lLBinning}, background noise suppression \cite{bkNoise}, power reduction~\cite{event}, and multi-resolution \cite{multires,frameMultires}, \etc~Since binning reduces the number of samples, it results in significant reduction in power consumption in an image sensor as ADC is responsible for major chunk of power consumption \cite{sony}. By decreasing the {\em bit resolution} of ADC one can reduce the consumption exponentially. This is because noise and linearity requirements are relaxed at smaller bit resolutions~\cite{adc}. In this paper, we use $2\times1$, $2\times2$ and $4\times4$ binning modes with averaging.

\subsection{Power and performance analysis of digital circuits} \label{Sec:pwrPerf}
Power consumption is one of the major concerns in digital circuits especially in mobile devices. The are two main independent components of power consumption/dissipation in a digital circuit - {\em Static power and Dynamic power}. Hence total power consumption can be written as 
\begin{equation}
P_{total} = P_{static}+P_{dynamic}.
\end{equation}
{\em Static power} is the power consumed when there is no activity in a digital circuit and {\em Dynamic power} is the power consumed due to switching signals in a digital circuit. Dynamic power can be further broken down as 
\begin{equation}
P_{dynamic} = P_{switching} + P_{short~circuit},
\end{equation}
where $P_{switching}$ refers to the power required to charge or discharge the switching nodes in digital circuit and $P_{short~circuit}$ refers to transient power consumption due to short circuit current when a gate switches from one state to another.  
\begin{equation}
P_{switching}= \alpha CV^2F,  \label{Eq:cv}
\end{equation}
where $\alpha$ is switching activity factor \ie, number of times a signal switches from 0 to 1 per cycle, $F$ denotes the frequency, $V$ represents the voltage and $C$ denotes switched capacitance at the node. $P_{switching}$ is one of the major concerns when comes to power consumption in a digital circuit. One can see from Eq. \eqref{Eq:cv} that power consumption is linearly proportional to frequency $F$ and quadratically proportional to voltage $V$. In general frequency of digital system is determined by the data processing requirements \ie, how much data must be processed per second. Thus if one wants to process half the amount of data in the same time, one can halve the operating frequency. Voltage and Frequency in Eq.~\eqref{Eq:cv} are not independent quantities but follow a proportional relationship. If the operating frequency is increased, the operating voltage must be increased and vice-versa due to device physics and noise margin requirements. 
Therefore, if there is a reduction in the data to be processed one can decrease voltage and frequency to achieve quadratic reduction in energy consumption. When this reduction in voltage and frequency is performed on the fly depending upon the data processing requirement it is called DVFS (Dynamic Voltage Frequency Scaling).

Another important factor when designing a digital system is {\em bitwidth} of the data. If the datapath is serial then it will imply longer time to transfer data and if it is parallel it will imply a wider data bus. 
Longer bitwidth implies larger arithmetic circuits such as adders, multipliers \etc or multiple clock cycles of operation. For instance, a single 4 bit adder can add two 8 bit numbers in 2 cycles or two 4 bit adder can add 8 bit numbers in one cycle.  This increases the delay of the most critical path or the slowest path in the circuit resulting in slower operation. To speed up one might have to use faster and power hungry circuits.

While there are many other factors impacting power and performance, a detailed analysis of those are beyond the scope of this work and only relevant issues are discussed here.

\subsection{JPEG} \label{Sec:JPEG}
After the sensor captures the image, the image is compressed by some method.
JPEG (Joint Photographics Expert Group) is one of the most widely used  lossy compression technique for images. Though JPEG can perform both lossy and lossless compression, the former one is popular due to little loss of perceptual quality. This is because most of the image information is contained in a very few coefficients in the discrete cosine transform (DCT) domain and hence insignificant coefficients can be discarded without much loss in perceptual quality producing large compression ratios. One can also control the amount of loss (and hence compression) by using the 'Quality' parameter of JPEG. It ranges from $1-100$ with 100 being the best lossy compression one can achieve. JPEG generally performs DCT on a block of $8\times8$ pixels, followed by quantization of DCT coefficients which constitutes the lossy step. After quantization codec is performed such as Huffman coding, run length coding \etc DCT is generally the most energy consuming part in JPEG compression \cite{jpeg}. Thus reduction in image data will lead to reduction in energy consumed in JPEG compression by a similar factor.

\subsection{Consideration in transmission and storage} \label{Sec:tranNstorage}
\begin{figure}[htb]
	\centering
	\centerline{\includegraphics[width=0.5\textwidth]{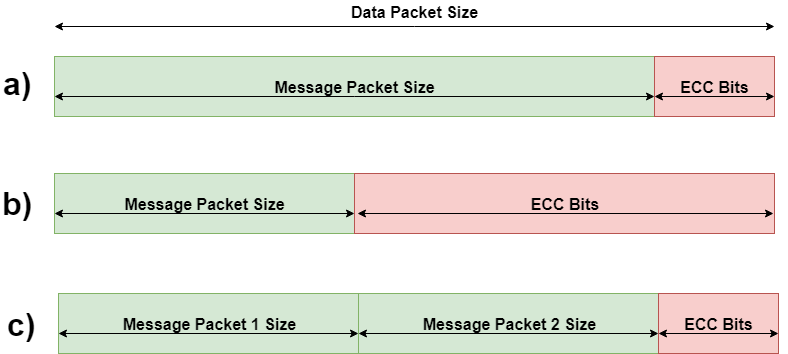}}
	\caption{Packet size analysis in wireless transmission.}
	\label{Fig:wt}
	\vspace{-3mm}
\end{figure}

After the image is compressed, \eg, via JPEG, into bit streams,
in wireless transmission, the data is encoded using ECC (Error Correction Coding) schemes to tolerate the errors occurring in wireless transmission due to channel noise. ECC schemes provide some type of redundancy in data to detect and correct the errors. One of the most popular ECC is LDPC (Low Density Parity Check Codes). It is used in both storage (\eg Solid State Drives \cite{ssd}) and wireless transmission (\eg 5G specifications \cite{3gpp}). LDPC uses parity check bits to detect and correct errors. A simple schematic of transmitted data packet is shown in Fig. \ref{Fig:wt}. It consists of message packet plus ECC bits (Fig. \ref{Fig:wt}(a)). If the channel is noisy then the transmitter would require to encode the message more strongly (Fig. \ref{Fig:wt}(b)). Thus for a fixed transmitted data packet size, actual message packet would become smaller resulting in reduced message bandwidth. If there is a possibility of compressing the message data then, one can either send more message bits per data packet resulting in increase of message transmission bandwidth or one can encode message bits more aggressively to make it more resistant to channel noise while keeping the message transmission bandwidth the same (Fig. \ref{Fig:wt}(b) and (c)). For storage the analysis it is analogous to that of wireless channel.  

\subsection{Bicubic interpolation} \label{Sec:Bicubic}
After the user receives the data and performs decoding and JPEG decompression, there is often a desire to obtain a high-resolution image based on the received low-resolution image captured by the sensor. 
Bicubic interpolation is widely employed as a baseline to perform super-resolution, which is an extension of cubic interpolation method. It considers a rectangular grid of $4\times4$ pixels and tries to fit a third order polynomial surface, $p(x,y)$, as follows - 
\begin{equation}
p(x,y) = \sum_{r=0}^{r=3}\sum_{c=0}^{c=3}a_{rc}x^ry^c,
\end{equation}
where $a_{rc}$ are the coefficients to be determined.
Bicubic interpolation  produces smooth images and is a popular benchmark for evaluating the performance of super-resolution algorithms \cite{ntire}.

\subsection{Deep residual networks}
\label{Sec:ResNet}
\begin{figure}[htb]
	\centering
	\vspace{-3mm}
	{\includegraphics[width=4cm]{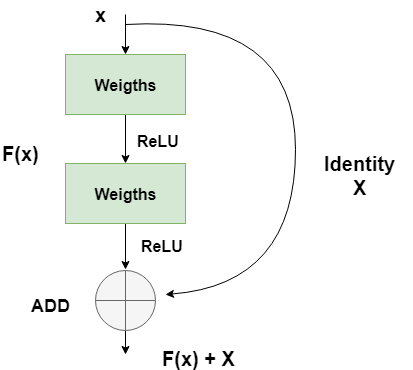}}
	\caption{Residual Network \cite{resnet}}
	\label{Fig:resnet}
	\vspace{-3mm}
\end{figure}

With the recent advances of ML, deep learning based algorithms have demonstrated superior performance than conventional methods for image super-resolution. Most of them are based on 
convolution neural networks (CNNs), which apply convolution operation to the input data followed by an activation function to produce the output.  To improve training of CNN, Residual Neural Network (ResNet) were first introduced by He \etal. in \cite{resnet}. To understand ResNets, let us denote the underlying mapping between input ($x$) and output of network as $H(x)$. Then residual mapping can be defined as, 
\begin{equation}
F(x) = H(x)-x.
\end{equation}
Simply speaking, residual mapping is the difference between input and expected output of network. The original mapping can now be defined in terms of residual mapping as 
\begin{equation}
H(x) = F(x)+x.
\end{equation}
A simple graph of residual network is shown in Fig. \ref{Fig:resnet}. ResNets performs superior because it is easier to optimize the residual mapping than the original \cite{resnet}. There is ample evidence in the literature  indicating that network depth is of crucial importance and deeper networks in general achieve better results \cite{verydeep,inception}. With ResNets it becomes easier to train big networks.

\section{Proposed acquisition and restoration methodology under hardware constraints.}
Bearing the above hardware constraints in mind, in this paper, we propose a new framework for image sensing and restoration using deep learning based technique to reconstruct the desired image.

\subsection{HCAS}
As mentioned before, a simple schematic of proposed sensing/acquisition methodology using HCAS (Hardware based Compressed Acquisition Scheme) is shown in Fig. \ref{Fig:methodology}. In this work, simulations using clean images is performed to mimic the proposed sensing methodology to verify the concept. It is expected that reconstruction performance will remain approximately the same in a real system.  

Entire imaging pipeline consists of 6 stages of which HCAS constitutes 3 stages. In the stage 1-3, image gets compressed using downsampling, bit truncation and JPEG. Stage 1-2 can be performed on the image sensor itself. Stage 3 happens on a JPEG chip or Digital Signal Processor (DSP). Stage 4 represents transmission of image which can be a wireless medium or an on-chip bus or even storage. Stage 5 can happen on a DSP processor in the device itself or it can be clubbed together with Stage 6 and can happen in the cloud or on the ML processor on the image acquisition device itself. The idea of this work is to save energy during acquisition \ie, from stage 1 to stage 5 as these processes often consume a significant amount of power in edge devices and stage 6 is not required unless a user is viewing image (\eg smart phones, surveillance cameras \etc) or a computer program is operating on images \eg object recognition. For some applications like drone transmitting a surveillance footage to a base station power consumption in stage 6 is not an issue.  
For stage 6, we propose DRCAS for image reconstruction. The subsequent paragraphs describe each process in more detail.

In stage 1, image is downsampled using simple binning (averaging operation) of pixels. The number of pixels that are averaged depends on downscaling factor. As mentioned before, while this operation happens on raw image, this work simulates this process on clean images for the sake of simplicity. Further lossy compression is performed using bit truncation in stage 2. We perform the task of bit truncation in the following way
\begin{equation}
Truncated~Pixel = round((averaged~pixel)\times2^{-N}), \nonumber
\end{equation}
where $N$ is the number of bits to be truncated. In this work $N$ is in the range $[0,3]$. 
Since the truncated pixel does not get multiplied with $2^N$ after rounding operation, the image appears darker after truncation operation for $N>0$. Simply speaking, we left-shift the pixel bits by the amount we want to truncate which makes the image appear dark. Thus the swing of the pixel values also gets reduced by a factor of $2^N$. This makes the image more compressible using JPEG in stage 3 because there is more loss of LSB bits than MSB bits in JPEG compression. This also means that there is more lossy compression induced artefact. The JPEG quality is varied in the range [70,100] in our experiments. Currently, we assume perfect transmission of image in stage 4. When JPEG image is decompressed in stage 5, it gets multiplied by a factor of $2^N$ to restore the brightness. Because of bit truncation and lossy JPEG compression, artefact get introduced in JPEG decompressed image in addition to the loss of resolution (caused by downsampling via binning).
Following this, our proposed DRCAS is employed to restore the desired high resolution image.
\begin{figure}[htb]
	\centering
	\vspace{-3mm}
	\centerline{\includegraphics[width=0.4\textwidth]{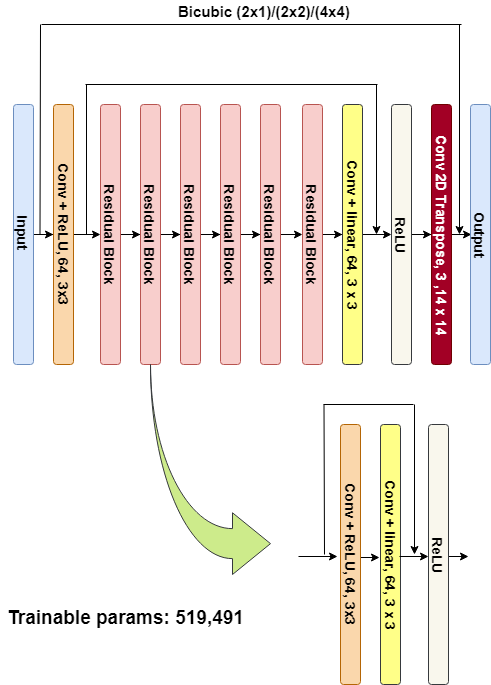}}
	\caption{DRCAS network proposed in this work.}
	\label{Fig:dnn}
	\vspace{-3mm}
\end{figure}
\begin{figure}[htb]
	\centerline{\includegraphics[width=0.45\textwidth]{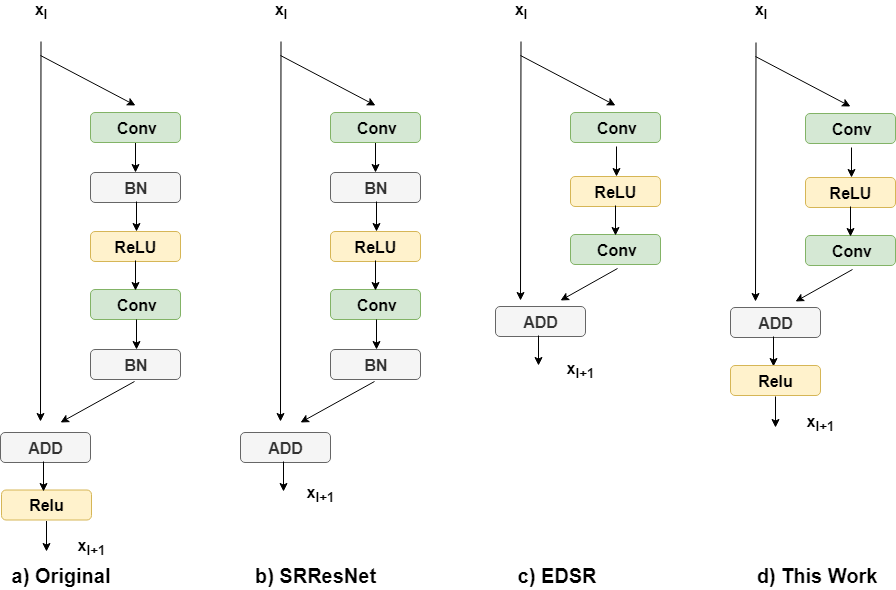}}
	\caption{A comparison of basic ResNet blocks}
	\label{Fig:resnetComp}
	\vspace{-3mm}
\end{figure}

\subsection{DRCAS}
We propose the DRCAS, denoting Deep Restoration network for hardware based Compressed Acquisition Scheme, to finish the task in Stage 6, with architecture shown in Fig.~\ref{Fig:dnn}. DRCAS is inspired by the original ResNet architecture \cite{resnet}, EDSR architecture \cite{edsr} and SRCNN architecture \cite{srgan} with some differences. A comparison of ResNet block is shown in Fig. \ref{Fig:resnetComp}. To start, the basic ResNet block used in this work uses ReLU (Rectified Linear Unit) layer in the end like original ResNet network. However it gets rid of Batch Normalization network as in EDSR Network. Also, unlike EDSR, our DRCAS avoids learning a complete image; it only learns the residual between the bicubic interpolated image and the actual image making the model much smaller in comparison. This is achieved by making a bypass connection between input and output using bicubic interpolation function as shown in Fig. \ref{Fig:dnn}. Since the residuals are mostly close to zero, the training is speeded up and the model complexity gets significantly reduced.

We train a separate network for a given downsampling factor, bit truncation and JPEG Quality factor to restore the image quality and resolution. Thus there are 48 different training tasks (4 cases of JPEG Quality, 4 cases of bit truncation and 3 cases of downsampling). The hyper-parameters are kept same for each training task. 

\begin{table*}[ht]	
	\begin{center}       
		\caption{Reconstruction results for DIV2K dataset (0701.png-0800.png). PSNR metric in dB. Bicub. refers to bicubic interpolation, EDSR refers to the EDSR network in paper \cite{edsr} and B.W. referes to the bitwidth of the image. } 
		\centering
		\resizebox{.8\textwidth}{!}
		{
		\begin{tabular}{|c|c||c|c|c||c|c|c||c|c|} 
			\hline			
			\rule[-1ex]{0pt}{3.5ex} $Quality$& $B.W.$&$This~Work$&$BiCub.$&$EDSR$&$This~Work$&$BiCub.$&$EDSR$&$This~Work$&$BiCub.$\\ 
			
			\rule[-1ex]{0pt}{3.5ex} &&$4\times4$&$4\times4$&$4\times4$& $2\times2$&$2\times2$&$2\times2$&$2\times1$&$2\times1$\\ \hline\hline
			
			\rule[-1ex]{0pt}{3.5ex}$100$	&	$8$ 	&	\boldmath{$27.91$}	&	$26.75$	&	$27.28$	&	\boldmath{$32.74$}	&	$30.97$	&	$31.61$ &	\boldmath{$34.96$}	&	$33.19$\\ 	\hline
			\rule[-1ex]{0pt}{3.5ex}$100$	&	$7$ 	&	\boldmath{$27.80$}	&	$26.68$ &	$27.12$	&\boldmath{$32.43$}	&	$30.78$	&	$31.14$ &\boldmath{$34.58$}	&	$32.86$\\ 	\hline
			\rule[-1ex]{0pt}{3.5ex}$100$	&	$6$ 	&\boldmath{$27.50$}	&	$26.44$ &	$26.59$	&\boldmath{$31.71$}	&	$30.09$	&	$30.04$ &\boldmath{$33.60$}	&	$31.96$\\ 	\hline
			\rule[-1ex]{0pt}{3.5ex}$100$	&	$5$ 	&\boldmath{$26.59$}	&	$25.72$	&	$25.26$	&\boldmath{$29.97$}	&	$28.76$	&	$27.87$ &\boldmath{$31.30$}	&	$29.98$\\ 	\hline\hline
			\rule[-1ex]{0pt}{3.5ex}$90$ 	&	$8$  	&\boldmath{$27.21$}	&	$26.36$	&	$26.29$	&\boldmath{$31.39$}	&	$30.18$	&	$29.88$	&\boldmath{$33.41$}	&	$32.12$ \\	\hline
			\rule[-1ex]{0pt}{3.5ex}$90$ 	&	$7$  	&\boldmath{$26.58$}	&	$25.87$	&	$25.51$	&\boldmath{$30.32$}	&	$29.35$	&	$28.69$	&\boldmath{$32.27$}	&	$31.11$ \\	\hline
			\rule[-1ex]{0pt}{3.5ex}$90$ 	&	$6$  	&\boldmath{$25.76$}	&	$25.12$	&	$24.63$	&\boldmath{$29.03$}	&	$28.14$	&	$27.38$	&\boldmath{$30.72$}	&	$29.64$ \\	\hline
			\rule[-1ex]{0pt}{3.5ex}$90$ 	&	$5$  	&\boldmath{	$24.61$}	&	$24.05$ &	$23.48$	&	\boldmath{$27.74$}	&	$26.49$	&	$25.75$	&	\boldmath{$28.62$}	&	$27.63$ \\	\hline\hline
			\rule[-1ex]{0pt}{3.5ex}$80$ 	&	$8$  	&	\boldmath{$26.60$}	&	$25.90$ &	$25.57$	&	\boldmath{$30.42$}	&	$29.44$	&	$28.82$	&	\boldmath{$32.37$}	&	$31.24$ \\	\hline
			\rule[-1ex]{0pt}{3.5ex}$80$ 	&	$7$  	&	\boldmath{$25.84$}	&	$25.23$	&	$24.80$	&	\boldmath{$29.23$}	&	$28.38$	&	$27.68$	&	\boldmath{$31.01$}	&	$29.98$ \\	\hline
			\rule[-1ex]{0pt}{3.5ex}$80$ 	&	$6$  	&	\boldmath{$24.93$}	&	$24.37$ &	$23.92$	&	\boldmath{$27.88$}	&	$27.06$	&	$26.43$	&	\boldmath{$29.40$}	&	$28.40$ \\	\hline
			\rule[-1ex]{0pt}{3.5ex}$80$ 	&	$5$  	&	\boldmath{$23.76$}	&	$23.20$	&	$22.69$	&	\boldmath{$26.13$}	&	$25.36$	&	$24.76$	&	\boldmath{$27.33$}	    &	$26.37$ \\	\hline
			\rule[-1ex]{0pt}{3.5ex}$70$ 	&	$8$  	&	\boldmath{$26.18$}	&	$25.55$ &	$25.15$	&	\boldmath{$29.76$}	&	$28.88$	&	$28.20$	&	\boldmath{$31.68$}	&	$30.60$ \\	\hline
			\rule[-1ex]{0pt}{3.5ex}$70$ 	&	$7$  	&	\boldmath{$25.39$}	&	$24.80$ &	$24.38$	&	\boldmath{$28.53$}	&	$27.73$	&	$27.09$	&	\boldmath{$30.25$}	&	$29.24$ \\	\hline
			\rule[-1ex]{0pt}{3.5ex}$70$ 	&	$6$  	&	\boldmath{$24.43$}	&	$23.87$ &	$23.44$	&	\boldmath{$27.13$}	&	$26.35$	&	$25.78$	&	\boldmath{$28.56$}	&	$27.58$ \\	\hline
			\rule[-1ex]{0pt}{3.5ex}$70$ 	&	$5$  	&	\boldmath{$23.16$}	&	$22.60$ &	$22.11$	&	\boldmath{$25.37$}	&	$24.58$	&	$24.03$	&	\boldmath{$26.52$}	&	$25.50$ \\	\hline			
		\end{tabular}
	}
		\label{Table:MLTable}
	\end{center}
\vspace{-6mm}
\end{table*}

\section{Experimental results}
We use DIV2K  dataset \cite{div2k} for training and evaluation; DIV2K dataset is a high-quality (2K resolution) image dataset, which consists of $800$ training images, $100$ validation images and $100$ testing images. Since the testing images are not made public, we use the last 100 images from training set (\ie~image 0701.png to 0800.png) as the testing set. The network uses patch size of $128 \times 128$ and $70,000$ training samples in a batch size of $64$. The network is trained for 24 epochs.

\begin{figure*}
	\begin{center}
		\begin{tabular}{ccc}
			\includegraphics[height=5cm]{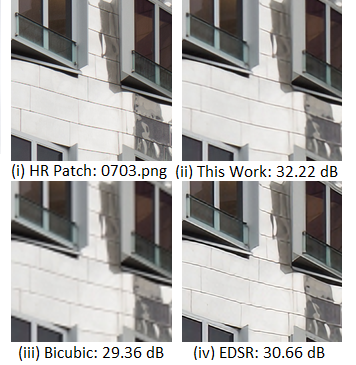}&
			\includegraphics[height=5cm]{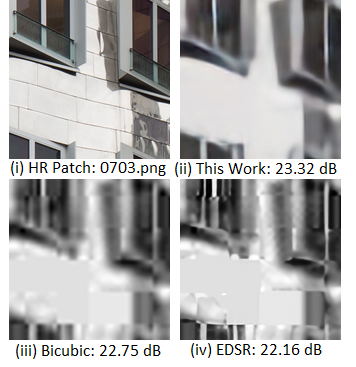}&
			\includegraphics[height=5cm]{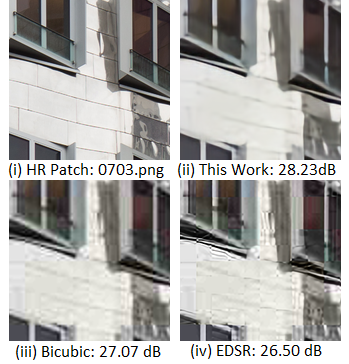}\\
			(a) $2\times2~S.R., Q = 100, B.W. = 8$  & (b) $ 4\times4~S.R., Q = 70, B.W. = 5$ & (c) $2\times2~S.R., Q = 80, B.W. = 6$\\
			\includegraphics[height=7cm]{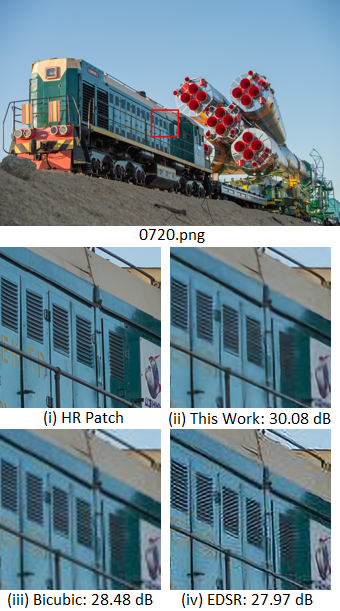}&
			\includegraphics[height=7cm]{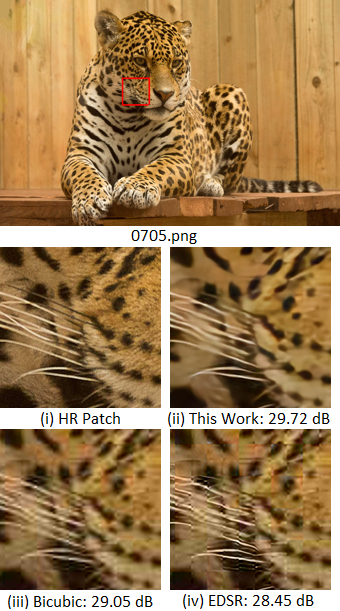}	&
			\includegraphics[height=7cm]{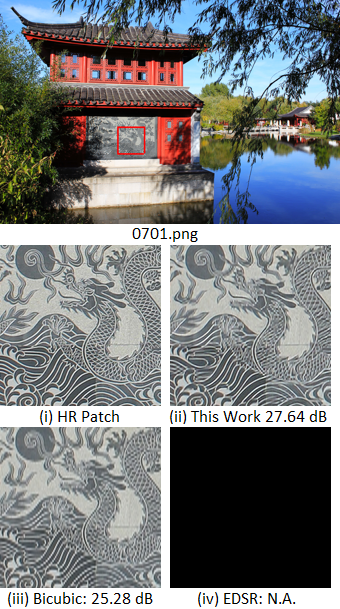}\\
			(d) $2\times2~S.R., Q = 90, B.W. = 7 $& (e) $ 2\times2~S.R., Q = 80, B.W. = 6 $& (f) $ 2\times1~S.R., Q = 100, B.W. = 8$\\
			\includegraphics[height=3cm]{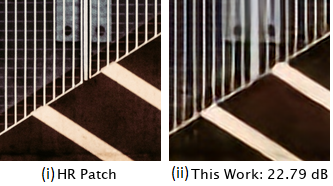}
			&\includegraphics[height=3.0cm]{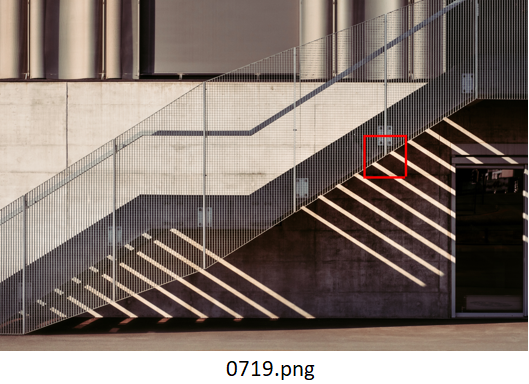}&
			\includegraphics[height=3cm]{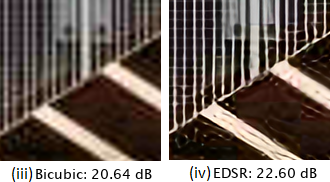}\\	
			&(g) $ 4\times4~S.R., Q = 90, B.W. = 6$ &
		\end{tabular}
	\end{center}
	\caption{Selected reconstructed images. For Fig. (f) EDSR result is not available as it is not designed for $2\times1$ super resolution.}
	\vspace{-5mm}	
	\label{Fig:sample}
\end{figure*}
Peak Signal to Noise Ratio (PSNR) is used as a metric to measure reconstruction performance with original high resolution image as the baseline. The results for reconstruction are shown in Table \ref{Table:MLTable} for full sized testing dataset. One can see from the result that our DRCAS outperforms bicubic. For testing purposes, this work also predicted the output of EDSR network for the downsampled images used in our work. As expected the performance of the EDSR network falls sharply as it is not trained to handle the noise due to bit truncation, averaging and JPEG. One can also see that the performance of EDSR gets worse than bicubic as the image quality degrades. It also serves as a proof that our DRCAS does train itself properly to handle the degradation induced by JPEG, bit truncation and binning. Some samples of reconstructed images including best case (Q = 100 and B.W. = 8) and worst case (Q = 70 and B.W. = 5) are shown in Fig. \ref{Fig:sample}. One can see that bicubic interpolated images are more noisy and less sharper than the images generated by DRCAS and EDSR.

\begin{table}[ht]
	\begin{center}      
		\caption{Comparison of networks.} 
		\centering
		\resizebox{.4\textwidth}{!}
		{
		\begin{tabular}{|c|c|c|} 
			\hline
			\rule[-1ex]{0pt}{3.5ex} Network& Residual~Blocks& Trainable~Weights\\ 
			\hline
			\rule[-1ex]{0pt}{3.5ex}  This~Work &$6$&$.5M$\\ \hline
			\rule[-1ex]{0pt}{3.5ex}  EDSR &$32$&$43M$\\ \hline	
		\end{tabular}
	}
		\label{Table:compEDSR}
	\end{center}
\vspace{-4mm}
\end{table}

While the performance of the network proposed in this work might seems inferior to numbers reported in the EDSR paper \cite{edsr}, the focus in this work is to {\em perform on-sensor compression to reduce data traffic and save energy}. This is achieved by bit truncation and pixel binning both of which can be performed on commercially available image sensors. Almost all existing works of super-resolution use bicubic downsampling method which yields better image but cannot be performed on the image sensor. Thus bicubic downsampling fails to perform compression of raw data. Our proposed DRCAS is also simpler than EDSR network and a comparison is shown in Table \ref{Table:compEDSR}.   Pixel binning and bit truncation lead to significant reduction in raw data generated from image sensor. An analysis of compression of raw data is provided in Table \ref{Table:RawTable}. It is measured as follows
\begin{equation}
Raw~Data~Compression = \frac{8\cdot N-B}{8\cdot N},
\end{equation}
where $N$ represents number of pixels binned together and $B$ represents bitwidth (B.W.) of pixel of downsampled image. One can achieve $50\% - 96\%$ reduction in raw data. Reduction in raw data means significant energy saving in downstream processing. One can achieve approximately proportional savings in energy for the same frequency of operation or one can employ DVFS to achieve quadratic scaling in energy reduction. Apart from savings in power, the system also becomes faster as there is less data to process.  Reduction in bitwidth can also result in exponential reduction in power consumed at ADC (Section \ref{Sec:imsensor}). Additionally, it can lead to more than proportional savings in energy in image processing circuits as LSB's switch more from one pixel to another than MSB in an image. 

\begin{table}[ht]
	\begin{center}       
		\caption{Raw data Compression Results. B.W. refers to the bitwidth of image. Raw Compression does not depend on JPEG Quality factor Q.}
		\centering
		\resizebox{.4\textwidth}{!}
		{
		\begin{tabular}{|c|c|c|c|c|} 
			\hline
			\rule[-1ex]{0pt}{3.5ex} & B.W. = 8&B.W. = 7&B.W. = 6&B.W. = 5  \\ 
			\hline
			\rule[-1ex]{0pt}{3.5ex}  $2\times1$ &$ 50\% $&$ 56.25\% $&$62.5\%$&$68.75\%$\\ \hline
			\rule[-1ex]{0pt}{3.5ex}  $2\times2$ &$ 75\% $&$ 78.12\% $&$81.25\%$&$84.37\%$\\ \hline
			\rule[-1ex]{0pt}{3.5ex}  $4\times4$ &$ 93.75\% $&$ 94.53\% $&$95.31\%$&$96.09\%$\\ \hline
		\end{tabular}
	}
		\label{Table:RawTable}
	\end{center}
\vspace{-4mm}
\end{table}

\begin{table}[ht]
	\begin{center}       
		\caption{Switching activity analysis of images. For DIV2K dataset (0701.png-0800.png). } 
		\centering
		\resizebox{.35\textwidth}{!}
		{
		\begin{tabular}{|c|c|} 
			\hline
			\rule[-1ex]{0pt}{3.5ex} $Bit~Position$& $Swiching~Activity~(\alpha$)\\ 
			\hline
			\rule[-1ex]{0pt}{3.5ex}  $0=LSB$ &$ 0.48$\\ \hline
			\rule[-1ex]{0pt}{3.5ex}  $1$ &$ 0.46$\\ \hline
			\rule[-1ex]{0pt}{3.5ex}  $2$ &$ 0.41$\\ \hline
			\rule[-1ex]{0pt}{3.5ex}  $3$ &$ 0.33$\\ \hline
			\rule[-1ex]{0pt}{3.5ex}  $4$ &$ 0.25$\\ \hline
			\rule[-1ex]{0pt}{3.5ex}  $5$ &$ 0.17$\\ \hline
			\rule[-1ex]{0pt}{3.5ex}  $6$ &$ 0.09$\\ \hline
			\rule[-1ex]{0pt}{3.5ex}  $7=MSB$ &$ 0.04$\\ \hline			                          
		\end{tabular}
	}
		\label{Table:SwitchingTable}
	\end{center}
\vspace{-6mm}
\end{table}

Let us assume that image is being readout to an 8-bit wide data bus in column-wise fashion for each color channel. A table of switching activity measurement for such a case is shown in Table \ref{Table:SwitchingTable}. This will lead to significant savings in dynamic power consumption as explained earlier in Sec. \ref{Sec:pwrPerf}.  Reduction in raw data also leads to reduction in processed image size after JPEG compression. The results for this are shown in Table \ref{Table:SizeTable}, which shows the size of the resulting image as a percentage of the size of the original high resolution image stored in lossless JPEG format. One can see that compressed image size ranges from $22.7\%$ to less than $1\%$ of the size of lossless image.

\begin{table}[tbp!]
	\begin{center}      
		\caption{Size Comparison. For DIV2K dataset (0701.png-0800.png). Measured as percentage with respect to original image in lossless JPEG format.}  
		\centering
		\resizebox{.4\textwidth}{!}
		{
		\begin{tabular}{|c|c|c|c|c|} 
			\hline			
			\rule[-1ex]{0pt}{3.5ex} $Quality$&$Bitwidth$& $2\times1$&$2\times2$&$4\times4$\\ \hline
			\rule[-1ex]{0pt}{3.5ex} $100$&$8$& $22.7\%$&$12.27\%$&$3.48\%$\\ \hline
			\rule[-1ex]{0pt}{3.5ex} $100$&$7$& $17.38\%$&$9.41\%$&$2.66\%$\\ \hline
			\rule[-1ex]{0pt}{3.5ex} $100$&$6$& $12.88\%$&$6.95\%$&$1.92\%$\\ \hline
			\rule[-1ex]{0pt}{3.5ex} $100$&$5$& $9.41\%$&$5.11\%$&$1.41\%$\\ \hline
			\rule[-1ex]{0pt}{3.5ex} $90$&$8$& $7.77\%$&$4.29\%$&$1.21\%$\\ \hline
			\rule[-1ex]{0pt}{3.5ex} $90$&$7$& $5.32\%$&$3.07\%$&$0.84\%$\\ \hline
			\rule[-1ex]{0pt}{3.5ex} $90$&$6$& $3.68\%$&$1.92\%$&$0.57\%$\\ \hline
			\rule[-1ex]{0pt}{3.5ex} $90$&$5$& $2.45\%$&$1.27\%$&$0.39\%$\\ \hline
			\rule[-1ex]{0pt}{3.5ex} $80$&$8$& $5.32\%$&$2.86\%$&$0.82\%$\\ \hline
			\rule[-1ex]{0pt}{3.5ex} $80$&$7$& $3.48\%$&$1.88\%$&$0.55\%$\\ \hline
			\rule[-1ex]{0pt}{3.5ex} $80$&$6$& $2.25\%$&$1.23\%$&$0.37\%$\\ \hline
			\rule[-1ex]{0pt}{3.5ex} $80$&$5$& $0.78\%$&$0.80\%$&$0.25\%$\\ \hline
			\rule[-1ex]{0pt}{3.5ex} $70$&$8$& $4.09\%$&$2.25\%$&$0.65\%$\\ \hline
			\rule[-1ex]{0pt}{3.5ex} $70$&$7$& $2.86\%$&$1.47\%$&$0.43\%$\\ \hline
			\rule[-1ex]{0pt}{3.5ex} $70$&$6$& $1.76\%$&$0.96\%$&$0.29\%$\\ \hline
			\rule[-1ex]{0pt}{3.5ex} $70$&$5$& $1.17\%$&$0.61\%$&$0.18\%$\\ \hline			
		\end{tabular}
	}
		\label{Table:SizeTable}
		\vspace{-7mm}
	\end{center}
\end{table}
As mentioned before, this work does not take into account the energy spent in reconstruction of the image as the aim is to reduce the energy for acquisition. The images can be reconstructed either on edge devices or on cloud, and an example of edge device can be SmartTV. One can reduce the image/video data transmission bandwidth. The image/video can then be reconstructed using the GPU available in smart TVs. For the case of smartphones, image can be acquired in low resolution mode and can be reconstructed back in the cloud. Since the images and photos are generally uploaded to cloud storage nowadays, reconstruction in cloud is technically feasible. For the case of drones transmitting video surveillance footage, the compression can reduce the power consumption in drone. It can also make transmission feasible in noisy environment or over longer distance because small image size offers an opportunity to aggressively encode the message packet with ECC.

\section{Conclusion and future work}
This paper establishes the use of DNNs for energy savings in process of image acquisition considering hardware constraints. For a given image quality requirement, one can acquire a low resolution image to save power and augment resolution and quality using DNN network in a fashion discussed in this paper. This would reduce the effort spent on the process of image acquisition resulting in improvement in power and performance parameters of the imaging device. The proposed methodology makes the system programmable \ie, user can shift from low resolution image acquisition to traditional high resolution image acquisition through software control of binning operation in imaging systems which is generally exposed to the system programmer. Using these techniques, one can achieve more than $50\%$ reduction in raw data and at least similar reduction in power while maintaining the PSNR above 30 dB. 

In the future, we would like to use a more accurate simulation for an imaging system. Deeper networks can also be studied to improve the reconstruction performance. We would also like to explore the prediction network using integer operation instead of floating point to make it suitable for edge devices and real time operation.

{\small
\bibliographystyle{ieee_fullname}
\bibliography{DRCAS_v1.bbl}
}

\end{document}